\documentclass[12pt]{article}
\usepackage{amsmath,amssymb,graphicx,mathrsfs,hyperref,slashed,color}
\newcommand{\be}{\begin{equation}}
\newcommand{\ee}{\end{equation}}
\newcommand{\bea}{\begin{eqnarray}}
\newcommand{\eea}{\end{eqnarray}}

\newcommand{\BH}{{\mbox{\tiny BH}}}
\newcommand{\mrm}{\mathrm}

\def\({\left(} \def\){\right)}

\renewcommand{\baselinestretch}{1.25}
\begin{document}
\title{\vspace{-1.8in}
{Lower limit on the entropy of black holes as inferred from gravitational wave
  observations }}
\author{\large Ram Brustein${}^{(1)}$,  A.J.M. Medved${}^{(2,3)}$, K. Yagi${}^{(4)}$
\\
\vspace{-.5in} \hspace{-1.5in} \vbox{
 \begin{flushleft}
  $^{\textrm{\normalsize
(1)\ Department of Physics, Ben-Gurion University,
    Beer-Sheva 84105, Israel}}$
$^{\textrm{\normalsize (2)\ Department of Physics \& Electronics, Rhodes University,
  Grahamstown 6140, South Africa}}$
$^{\textrm{\normalsize (3)\ National Institute for Theoretical Physics (NITheP), Western Cape 7602,
South Africa}}$
$^{\textrm{\normalsize (4)\ Department of Physics, University of Virginia, Charlottesville, Virginia 22904, USA}}$
\\ \small \hspace{1.07in}
    ramyb@bgu.ac.il,\  j.medved@ru.ac.za,\ ky5t@virginia.edu
\end{flushleft}
}}
\date{}
\maketitle
\begin{abstract}

Black hole (BH) thermodynamics was established by Bekenstein and  Hawking, who made abstract theoretical arguments about the second law of thermodynamics and quantum theory in curved spacetime respectively. Testing these ideas experimentally  has, so far, been impractical because the putative flux of Hawking radiation from astrophysical BHs is too small to be distinguished from the rest of the hot environment. Here, it is proposed that the spectrum of emitted gravitational waves (GWs) after the merger of two BHs, in particular the spectrum of GW150914, can be used to infer a  lower limit on the magnitude of the entropy of the post-merger BH.  This lower bound is potentially significant as it could be of the same order as the Bekenstein-Hawking entropy. To infer this limit, we first assume that the result of the merger is an ultracompact object with an external geometry which  is Schwarzschild or Kerr, but with an  outer surface which  is capable of reflecting in-falling GWs rather than fully absorbing them. If the absence of deviations from the predictions of general relativity in detected GW signals will be verified, we will then obtain a bound on the minimal redshift factor of GWs that emerge from the vicinity of the object's surface. This lack of deviations would also mean that the remnant of the merger has to have a strongly absorbing surface and  must then be a BH for all practical purposes. We conclude that a relationship between the minimal redshift factor and the BH entropy, which was first proposed by 't Hooft, could then be used to set a lower bound on the entropy of the post-merger BH.
\end{abstract}
\newpage
\renewcommand{\baselinestretch}{1.5}\normalsize

\section{Introduction}

It has long been thought that the picture of a black hole (BH) from classical general relativity (GR) accurately  depicts the end point of gravitational collapse, perhaps with a suitable regularization scheme for taming the infamous BH singularity. This belief has recently been questioned due to the realization that, even as a matter of principle,  the paradoxical nature of BH evaporation has no simple remedy along the lines of observer complementarity
\cite{Sunny,Mathur1,Braun,Mathur2,AMPS,MP,MathurCP}.
One then concludes that
the synthesis of classical BHs with the quantum process
of Hawking radiation is untenable as currently understood.
A self-consistent picture requires discarding one, at the very least, of several
cherished tenets: unitary evolution, locality, causality,
the strong subadditivity of entropy (or monogamy of entanglement),  Einstein's principle of  equivalence  and/or effective field theory in its domain of applicability.

A different route is to assert that the traditional model of a BH is what needs to be changed and then  argue on behalf of an exotic alternative. This idea has led to models for ultracompact objects that resemble classical BHs in some ways yet  differ in others.  Proposals of this nature include fuzzballs \cite{Mathur1,otherfuzzball}, firewalls \cite{Braun,AMPS}, wormholes \cite{Cardoso}, gravastars \cite{MM}, anti-de Sitter bubbles \cite{bubble}, boson stars \cite{bozo}, graviton condensates \cite{Dvali}, strongly anisotropic stars~\cite{Bowers:1974tgi,Yagi:2015upa,Yagi:2016ejg,Raposo:2018rjn} and, as promoted by the current authors, collapsed polymers \cite{strungout}. A partial catalogue of sorts was recently  presented in \cite{Cardosorev}. To avoid the paradoxical dilemmas of an evaporating BH,  an ultracompact object  cannot have a  ``true''  horizon, although its outer surface may still display some horizon-like characteristics. As such, it lacks a surface where the Tolman redshift factor, $\sqrt{|g_{tt}|}$  vanishes locally:  a surface of infinite blueshift. The Tolman redshift factor is related to the
standard redshift parameter $z$ as  $\;\sqrt{|g_{tt}|}  = (1+z)^{-1}\;$. The
minimal value of  the Tolman factor or, equivalently, the maximal  value  of $z$ is an important parameter that may be used to distinguish among the different   models.

The BH entropy, on the other hand, can be interpreted as a geometric quantity or as a measure of entanglement across the horizon. The entropy has, at times, also been attributed to the number of  internal microstates, the amount of information that is shielded by the horizon and so forth (see \cite{bek} for further discussion along these lines). We would argue that all these are different descriptions of the same entropy, as distinct observers could well have conflicting perspectives on its origin.

For an asymptotic observer who is aware of the existence of quantum fields, a small value for the Tolman redshift factor implies that the near-horizon  modes of these fields have very large energies and  correspondingly small wavelengths. As a result, the observer expects many such modes to be closely squeezed together in the proximity of the horizon and form a ``thermal atmosphere" around the BH horizon \cite{brick}. The density of states of the near-horizon modes increases as the Tolman redshift factor decreases and therefore, so does the entropy of the thermal atmosphere.  In the case of our asymptotic observer ---  who, as a disciple of GR, only knows about what is outside the horizon and would {\em assume} a mostly empty interior  --- the entropy of the BH would have to be attributed to this thermal atmosphere. The asymptotic observer would then conclude that the magnitude of the BH entropy depends on the value of the minimal Tolman redshift factor. But a free-falling observer, for example, is unaware of any redshift effect and would have to conclude differently.

Gravitational waves (GWs) that are emitted from BH mergers \cite{LIGO,TheLIGOScientific} provide a means for  constraining  deviations away from GR \cite{LIGO1,Konoplya:2016pmh,pretorius}. Our proposal is  that such  GW  data, which  can already  be used to put an upper bound on the observed minimal value of the Tolman factor, can similarly be used to infer a lower bound on the entropy of the post-merger  BH. This means that  BH mergers, which can be viewed as cosmological ``scattering experiments'',   can be  used to  infer the entropy of  astrophysical BHs. As far as we are aware, no similar proposal has been made in this context.

The  more exotic compact objects allow for (at least) two other additional sources of emitted  waves as compared to the BHs of GR.  First,   an  ultracompact object containing  a significant amount of matter could very well support modes that are analogous  to  the fluid modes of relativistic stars \cite{AKK}. For further discussion on these ``fluid-like'' modes, see \cite{spinny,collision}. Another class of waves to consider are  those that  are sourced  by the so-called echoes \cite{Cardoso,moreCardoso,noncardoso,Conklin:2017lwb,Price,Westerweck:2017hus,Cardosorev}, which were discussed for rotating compact objects in  \cite{Pani:2010jz,Cardoso:2014sna,Maggio:2017ivp,Barausse:2018vdb,yanbi,Maggio:2018ivz}. The echoes for a star-like ultracompact object are modes that reflect any number of times between the peak of the potential (or photosphere)  and the object's outermost surface, whereas  wormhole echoes \cite{moreCardoso} are reflected between one potential peak and its mirror image.
We note that a reflection from the outermost surface is essentially equivalent to unhindered propagation through the object, as the waves would then be perceived as being totally reflected at the center of the object.  In any event,  the reflected modes  will eventually escape from their respective cavity,  or ``echo chamber'',  and move  outward through the gravitational barrier. The reason that these other modes can  be distinguished (at least in principle) from the standard BH modes is because of an  additional length scale, $L_i$ for the $i^{th}$ exotic alternative. This scale will be made explicit in the section to follow.

The value of these additional length scale $L_i$ does not necessarily fix the precise type of ultracompact object. Some models are endowed with a natural choice of length scale but then some are not. For instance, in the case of the collapsed-polymer model, $L_i$ is related to the string length. The string length must be greater than Planck length $l_P$, but its precise value is  otherwise difficult to pinpoint. In comparison,  $L_i$ is a completely free parameter for the wormhole model whereas, for the gravastar model, it is related to the width of an outer shell which is supposed to be about twenty orders of magnitude larger than $l_P$ \cite{MM2}.

We do support the point of view that GR is only an approximate classical theory and there  should, as such, be  corrections to the spectra of both the ringdown modes and  the  emitted GWs which follow after a BH merger. On general grounds, and as argued in \cite{collision}, these excitations should have lower frequencies and longer lifetimes in comparison to the standard GR modes.  But, since we expect that the boundary conditions at the surface are, to a good approximation, totally absorbing boundary conditions, it follows that the amplitude of the additional modes has to be substantially suppressed as compared to the GR modes.
The same reasoning would  apply to the aforementioned fluid-like  modes. Hence,
for the purposes of the current discussion, GR-corrected and fluid-like modes can be discounted.

The echo modes, on the other hand,  represent the complete opposite of our expectations, as these obey totally reflecting  boundary conditions and serve as a  means to quantify the deviations away from GR in this extreme region of parameter space. As already noted, the simplest way to have reflecting boundary conditions is if, for some unknown reason, the GWs pass undisturbed through the interior of the BH (as in the gravastar model). Such waves would indeed be viewed from the exterior as being totally reflected. One could, of course, invent intermediate models; for example, by modifying the gravitational potential by hand.  It does, however, seem difficult to justify such impromptu modifications. In any event,  the resulting spectrum of emitted GWs would still be expected to be qualitatively similar to the case of total reflection, provided that the reflection is not parametrically small. This point is clarified later.

The rest of the paper starts by  establishing the quantitative relationship between the entropy  and the minimal redshift factor for the  BHs of GR  and then, similarly,  for the exotic alternatives. This  is followed by a numerical analysis that determines the current   bounds on the redshift factor and
then shows how the corresponding  bounds on the entropy could be inferred.
 The paper concludes with a brief overview.

\section{Entropy and redshift}

Here, we will  elaborate on the significance of the  redshift factor, with the
focus being initially  on the BHs of GR and, following this, on the   exotic alternatives.  We will, as a starting point,  assume non-rotating and electrically neutral compact objects in $3+1$ dimensions, so that  the object's exterior geometry  is accurately described by the Schwarzschild metric $g_{ab}$, for which $\;-g_{tt}=g^{rr}=1-\frac{R_S}{r}\;$ where  $R_S$ is  the Schwarzschild radius.  Rotating BHs  will, however,  be eventually  considered; in which case,  the exterior geometry is described by the Kerr metric with  $\;-g_{tt}=  1- \frac{2 M r}{r^2 + a^2 \cos^2\theta}\;$ and $\;g^{rr}= \frac{r^2-2Mr+a^2}{r^2 + a^2 \cos^2\theta}=\frac{(r-r_+)(r-r_-)}{r^2 + a^2 \cos^2\theta}\;$. Here, $M$ is the mass, $\;a= J/M\;$ is the dimensional spin parameter and  $\;r_{\pm}=M\pm M\sqrt{1-a^2/M^2}\;$  is the location of the outer/inner horizon (or would-be horizons for the alternative models).

\subsection{Minimal redshift for black holes and exotic alternatives}

Even in the idealized case of GR BHs, the notion of a vanishing redshift factor is not realistic. Smaller distances than $l_P$ are impossible to probe without turning the probing instrument  itself into a BH.
And so an observer cannot have better resolution than $l_P$ when attempting to locate a BH horizon.

The following calculation reveals that a proper distance $\Delta r$ of one Planck length  away from the horizon  describes a radial distance  in Schwarzschild coordinates of about $\;d= l_P^2/R_S\;$:
\begin{equation}
\label{propL}
\Delta r\;=\;\int^{R_S+d}_{R_S} dx \sqrt{g_{rr}(x)}\;\simeq\;\sqrt{d \, R_S}\;\;.
\end{equation}
This, in turn, implies that the  surface has a local (Tolman) redshift factor of
\be
\label{minredshift}
F\;\equiv\;\sqrt{|g_{tt}|}
\;\simeq\; \frac{\Delta r}{R_S}\;\geq\; \frac{l_P}{R_S}\;.
\ee
The GR value for the minimal redshift factor is then $\;F_{\BH}=l_P/R_S\;$. (See also \cite{RBKupf,origin,Brustein:2013xga} for a different perspective leading to the same conclusion.) For a solar mass BH,  the minimal redshift factor is extremely small,  $\;F_{\BH}=5\times 10^{-38}$.

The Bekenstein--Hawking value of the BH entropy, $\;S_{\BH}= A/4l_P^2 =\pi R_S^2/l_P^2\;$, can now be expressed as $\;S_{\BH}= \pi F^{-2}_{\BH}\;$. As mentioned in the Introduction and explained in detail in the next subsection, this reasoning is
based on  ascribing the BH entropy to the thermal atmosphere of the closely surrounding quantum field modes.

\subsection{Entropy and minimal redshift}

The properties of the thermal atmosphere  were first calculated by 't~Hooft~\cite{brick}
by counting  the number $n(E)$ of  excited quantum field  modes
having  energy $E$ and satisfying the wave equation
$\;g^{rr} k^2 = -g^{tt} E^2+\cdots\;$.   This was done by using the Bohr quantization condition to determine $n(E)$,  $\;\pi n(E) = \int dr k(E) \;$, where $k(E)$ is the wave number. Since $\;k(E) \sim \sqrt{-g^{tt}g_{rr}} E \sim E/|g_{tt}|\;$,  it follows that $n(E)$, just like  $k(E)$,  is  inversely proportional to the square of the minimal redshift factor. Using this estimate, 't~Hooft proceeded to show that the free energy and therefore the energy and entropy grow in inverse proportion to the square of the minimal redshift factor. (See  \cite{RBKupf} for further  details.) Actually, the estimates of 't~Hooft are only  valid for cases  in which the minimal redshift factor $F_\mathrm{min}$ is parametrically small, $\;F_\mathrm{min} \ll \;1$. This limitation will be important later when we derive bounds on $F_\mathrm{min}$.

Since the entropy formally diverges when the minimal redshift factor vanishes, 't~Hooft regularized the result by placing a hard cutoff (a ``brick wall") at a small distance away from the horizon, which he later identifies as a proper length on the order of $l_P$. A specific choice of cutoff  reproduces the expected value of the BH entropy $S_{\BH}$; however, 't~Hooft could  just as well have  followed our prescription for a minimal redshift factor to  obtain the correct value.   This provides a formal explanation as to why a BH can have a parametrically much  larger entropy  than a run-in-the-mill star of the same
mass.~\footnote{Although BHs and compact stars can still
have a  similar entropy
if it is rescaled by the object's mass and temperature~\cite{Oppenheim:2002kx,Alexander:2018wxr}.} And there is no ``species problem'' in this regard because the renormalized  value of $l_P$ depends on the number of species ${\cal N}$ \cite{RBYarom,Dvali2,BDV}, and it does so in just the right way to cancel ${\cal N}$ out of the brick-wall calculation.

The case of rotating BHs was discussed in \cite{brickrot} and the result is very similar to that of the Schwarzschild case. The area of the BH horizon is now given by $A= 2\pi M r_+$ and the entropy is again inversely proportional to the square of the minimal redshift factor, $\;S\propto F^{-2}_{\BH}\;$.

Let us now consider the other ultracompact objects, the exotic alternatives. Each of these can be expected to have a characteristic length scale that represents the displacement of the object's outer surface from its Schwarzschild radius  and, with it, an associated minimal redshift factor.   Adopting and then slightly modifying a definition from \cite{Cardosorev}, we define $\;\epsilon_i=(\Delta r)_i/R_S\;$ as a dimensionless form of the characteristic length for the $i^{th}$ model. Since $\epsilon_i$  is also the near-horizon redshift factor, as made clear by  Eqs.~(\ref{propL}) and~(\ref{minredshift}), it follows that the object has a minimal redshift factor of $\;F_{i,\mathrm{min}}=\epsilon_i\;$ and then a thermal atmosphere entropy of  $\;S_i \propto 1/\epsilon_i^2\;$. We can absorb the constant of proportionality between $\;S_i$  and $1/\epsilon_i^2\;$ into the definition of $(\Delta r)_i$ and then write
\be
\label{entropy1}
S_i\; =\; 1/\epsilon_i^2\;,
\ee
which is the adopted  form of entropy for  the remainder of the paper. A related discussion in the context of BH alternatives can be found in \cite{Holdom:2016nek}.

\subsection{A bound on redshift and entropy}

The basic idea that enables  the bounding of the minimal redshift factor $\epsilon_{i,\mathrm{min}}$  will now be explained.  We begin this discussion with non-rotating BHs from  GR and  later move on to include rotation and/or alternative models.
In the classical-BH case, any emitted GWs can be attributed to   damped ringdown or quasinormal modes (QNMs)  which  ensue when a perturbed BH --- such as one formed from the merger of a BH binary ---  settles down to its equilibrium state.~\footnote{There are many excellent reviews on the QNMs of BHs; see, {\em e.g.}, \cite{QNMBH,QNMBH2,Konoplya:2011qq}.}  To understand the properties of these QNMs, one can consider  waves propagating in from infinity that  are scattered by  the gravitational potential barrier at $\;r\simeq \frac{3}{2}R_S\;$. There are then two possibilities: the wave is either reflected off the barrier or  passes right through it.  The first possibility describes the emitted  GWs, whereas the second class of waves are never heard from again because a classical BH is a perfect absorber.~\footnote{Quantum mechanically, a  minimal amount of reflection can be expected at a rate that is compliant with the Hawking rate of emission.} For future reference, the emitted waves have an angular frequency $\omega_{\BH}$ of about $\;R_S \omega_{\BH} \sim 1\;$ and the longest-living modes have an inverse damping time $\tau^{-1}_{\BH}$ of similar magnitude, $\;R_S \tau^{-1}_{\BH}\sim 1\;$.  These estimates reflect the fact that a Schwarzschild BH has only one length scale $R_S$. For the exact numerical values of $\omega_{\BH}$ and $\tau_{\BH}$, see  \cite{QNMBH,QNMBH2,Konoplya:2011qq}.

Moving on to the exotic alternatives, let us
recall that the other types of modes can  be distinguished  from the standard BH modes by an  additional length scale. Here, rather than the scale $\epsilon_i$,  it is more appropriate to consider the distance
$L_i$ from the object's outermost surface to its photosphere {\em but} in terms of the tortoise coordinate  $\;r_{\ast}=\int \frac{dr}{1-\frac{R_S}{r}}\;$.
The advantage of this coordinate choice is that  the near-horizon geometry then looks  similar to a flat geometry.
And so  the relevant length scale for the $i^{th}$ model is  given by
(the photosphere is placed at $\;r=(3/2) R_S\;$ for simplicity)
\be
L_i \;=\;
\int\limits^{R_S/2}_{(\Delta r)_i}\frac{d(r-R_S)}{1-\frac{R_S}{r}}
\;\simeq\; R_S\ln{\left[\frac{R_S}{2 (\Delta r)_i}\right]}
\;=\; R_S|\ln{\epsilon_i}|\;,
\ee
where we have assumed that $\epsilon_i$ is small. In the rotating case, $\;r_{\ast}=\int dr \frac{r^2+a^2}{r^2+a^2 - 2 Mr}\;$ and the expression for $L_i$ is similarly obtained, $\;L_i=\; r_+|\ln{\epsilon_i}|\;$, valid for
$\;\epsilon_i\ll 1\;$.

The frequency of an echo mode, in the case of a  non-rotating BH, is determined by the inverse of the total length of its trip ({\em i.e.}, the length of the cavity times the number of reflections). Hence, the highest-lying angular frequency of the echo modes would go as
\be
\label{eq:omegaI}
R_S \omega_i  \;\sim\; \frac{R_S}{L_i} \;=\;
\frac{1}{|\ln{\epsilon_i}|}\;.
\ee
As $\;\epsilon_i\ll 1\;$ for most models of interest,  the frequency of even the most rapidly oscillating modes of the echoes is always suppressed relative to
$\omega_{\BH}$. Note that we are using the notation $\omega_i$ to denote the real part of the frequency of a mode in the $i^{\rm th}$ model of exotic objects. Instead of the imaginary part of the frequency, we will use the lifetime $\tau_i$.

When the compact object  is rotating, the frequency in Eq.~(\ref{eq:omegaI})  is modified by the rotation in accordance with \cite{Starobinsky,Vilenkin} (also see \cite{Maggio:2018ivz})
\be
\label{eq:omega_R-maggio}
\omega_i \;=\; - \frac{\pi }{2 |r_*^0|}\left(q+\frac{s(s+1)}{2}\right)+ m\Omega\,,
\ee
where
\begin{align}
r_*^0 &\;\sim\; M [1+(1-a^2/M^2)^{-1/2}] |\ln \epsilon_i|\,, \\
\Omega &\;=\; \frac{a/M}{2 r_+}\;.
\end{align}
The positive integer $q$ in Eq.~(\ref{eq:omega_R-maggio}) is  denoting the overtone number, for which we will always choose the dominant one, $\;q=1\;$ and $s$ is denoting the spin of the mode, for which we will always choose $s=2$, corresponding to a GW. Because $\omega_i-m\Omega$ is negative for $\;\epsilon_i\ll 1\;$ and $\;a\lesssim M\;$, the typical solution is formally unstable, just like the r-mode solutions for neutron stars ({\em e.g.}, \cite{LOM}). However, the instability is very weak and we will, conservatively, ignore the possible growth of the echo modes due to this effect.

The inverse of the  damping  time is  also suppressed, but even more so because of the very small cost in wave amplitude for any given trip through the cavity (however, see the second caveat below). A calculation reveals that the inverted decay time for the most rapidly oscillating  mode  would be given by \cite{Cardosorev}
\begin{align}
\label{eq:tau_I-maggio}
1/\tau_i &\;\sim \;- \frac{2 \beta_{s\ell}M}{|r_*^0|} \frac{r_+}{r_+ - r_-}[\omega_i(r_+-r_-)]^{2\ell+1}(\omega_i-m\Omega)\,,
\end{align}
with  $\;s=2\;$  and $(\ell,m)$ representing the spin and  harmonic mode, respectively,  of the graviton, and $\beta_{s\ell}$ is a spin and mode-dependent number. The mode  will be chosen  as $(2,2)$ and, with this choice,  $\;\beta_{22} = 1/225\;$.

For the case of interest, $\;|\ln \epsilon_i|\gg 1\;$ and $\;s=\ell=m=2\;$,
\begin{align}
\label{eq:omega-app}
\omega_i &\;\simeq \; m\Omega\;=\;\frac{a/M}{r_+}
\end{align}
and
\begin{eqnarray}
\label{eq:tau_I-app}
\tau_i & \simeq & \frac{225}{ \pi}\frac{1}{M}  |r_*^0|^2
(a/M)^{-5} \left(\frac{r_+ - r_-}{ r_+}\right)^{-4}\cr
& = &  \frac{225}{ \pi} r_+ |\ln \epsilon_i|^2  \frac{1+\sqrt{1-a^2/M^2}}{1-a^2/M^2}(a/M)^{-5}
\left(\frac{r_+ - r_-}{ r_+}\right)^{-4} \;.
\end{eqnarray}
Notice that  the decay time of the echoes is fortuitously long because of the
factor of  $ |\ln \epsilon_i|^2$. This huge enhancement in the lifetime of the additional modes is essential for their detection.

To get a better idea about the magnitude of the numerical coefficients,  one can evaluate $\tau_i$ for the case $\;a=0.68 M\;$, which corresponds to the remnant BH of the famous detection GW150914 \cite{LIGO},
\be
\tau_i \;= \;3100\; r_+ |\ln\epsilon_i|^2\;.
\ee
So that, in addition to the enhancement due to the large value of $|\ln\epsilon_i|^2$, one finds an  equally large numerical coefficient.  Indeed, 3100 is  the same order of magnitude as what $|\ln\epsilon_i|^2$ would be for the inverse of the square root of the  Bekenstein--Hawking entropy, $\;\epsilon_{Bek-Hawk}=1/\sqrt{S_{\BH}}
\sim 10^{-40}\;$.

One can understand, heuristically, the scaling of  the lifetime of the echoes  in terms of ideas that were introduced in \cite{ridethewave}. To make the argument, we will assume that the gravitational force acts democratically and thus weakly with all forms of matter, standard-model based or otherwise. The matter itself can interact strongly and be quite exotic. We believe that this is a reasonable assumption.

First recall
from Eq.~(\ref{eq:omega_R-maggio}) that  the ``intrinsic" angular frequency is $\;\omega_{i} \sim \frac{1}{r_+}\frac{1}{|\ln{\epsilon_i|}}\;$
(which is further modified by the external rotation $m\Omega$; however, the intrinsic frequency is that seen by a co-rotating observer which is the appropriate one for the current discussion). This means that the GR external observer,  who is unaware of the origin of the echo modes, would conclude that the wavelength of the radiation at distances far away from the horizon is
$\;\lambda_i\sim r_+ |\ln{\epsilon_i}|\;$.  On the other hand, the same observer  attributes the source with having   an area of about $M r_+$. She then just needs to know that the transmission cross-section for such long wavelength modes through a proportionally smaller surface of area $A$ is determined by the ratio $A/\lambda^2$, which translates into $Mr_+/\lambda_i^2\sim\frac{1}{|\ln{\epsilon_i}|^2}\;$.
The lifetime is inversely proportional to the transmission rate and therefore scales as  $\;\tau_i\sim |\ln{\epsilon_i}|^2\;$. This estimate is based on the assumption that most of the mode's energy is being emitted in the form of coherent waves rather than  dissipating as heat (see the third caveat below).

The above heuristic argument can also be used to discuss a compact object whose reflection at the would-be horizon is not so perfect. Our contention is that, unless the reflection is parametrically small (in which case, the absorption is essentially complete and then the object effectively has a horizon), the  mode's lifetime scales in the same manner as it does for the case of total reflection. The key point  is that partial absorption occurs at the surface of the compact object. Then, given the above logic, the energy which is absorbed through that surface would similarly scale as $\;A/\lambda_i^2\sim\frac{1}{|\ln{\epsilon_i}|^2}\;$. But, for small $\epsilon_i$, the area of the object's surface scales the same as the area of
the  gravitational potential barrier.
It then follows that the energy absorption through the object's surface scales in the same way as its energy leakage through the potential barrier, given that the wavelengths of the leaking modes are similarly set by the minimal redshift (as argued in \cite{ridethewave}).
And, since it is the leakage through the barrier that determines the lifetime of the emitted waves,  the conclusion is that, up to some numerical factor, Eq.~(\ref{eq:tau_I-app}) applies just as well for  the case of a finite absorber.

We emphasize that the scaling $A/\lambda^2$ results from the assumption that the gravitational force acts democratically on all forms of matter, including whichever could be found inside exotic objects. We also would like to point out that the effects of strong blueshifting could only be parametrically  significant  inasmuch as the blueshifting is exponentially strong, otherwise $|\ln{\epsilon}|$ would be of order unity.  But, if the blueshifting was exponentially strong, then the object would be a black hole for all practical purposes and not an exotic alternative. The detectability of the echo modes depends strongly on the actual value of the absorption \cite{Testa:2018bzd}.

In principle, current observational data from the BH merger event which led to GW150914 \cite{LIGO,LIGO1} can be used to support an upper limit on the
minimal value of $\epsilon_i$ (see Section~3 for the precise statement and details),
\be
\epsilon_{i,\mathrm{min}} \;\lesssim \; 10^{-40} \;;
\ee
which can, according to our proposal, be used to bound the entropy from below,
\be
S\; = \;\frac{1}{\epsilon_{i,\mathrm{min}}^2} \;\gtrsim\;  10^{80}\;.
\ee
Although further scrutiny is still necessary, this bound provides the first  experimental indication that an astrophysical BH possesses extremely large entropy, on the order of that predicted by the Bekenstein--Hawking area law. And given that the BH has  both an entropy and  energy (or a  mass), a temperature
must then follow.

\subsection{Caveats}

Like in any complicated physical situation, the above picture has caveats. Let us touch upon just a few:

\begin{enumerate}
\item
For the BHs of  GR, the spectral properties of the QNMs depend strongly  on
both the external geometry and the boundary conditions at the horizon (those of in-falling waves only).
And so, whether or not there are spacetime modes akin
to the standard modes of a GR BH  will depend on the boundary conditions at the outermost surface of the model in question.
It is quite possible that the absence or modification of the conventional spacetime
modes is already enough to rule out some otherwise-plausible compact objects.  The exterior geometry of any non-rotating  compact object is expected to be a Schwarzschild geometry due to Birkhoff's theorem, if GR is assumed to be valid. On the other hand, the exterior geometry of a rotating object could differ noticeably from that of Kerr; but this correction is expected to be small  when $\;\epsilon_i\ll 1\;$  \cite{PPani}.
A notable exception is the polymer model because, as explained in \cite{collision}, it behaves just like  a perfect absorber in the limit that the string coupling (equivalently, $\hbar$) goes to zero.
Hence, the polymer BHs are emitters of both conventional spacetime modes and fluid modes; the latter of which exhibit some similarities with the scaling properties of the echo modes thanks to the previous barrier-leakage argument \cite{ridethewave}.

\item
It is sometimes assumed that  the echoes have a suppression in amplitude, relative to the  initial burst, which is  in addition to that of   Eq.~(\ref{eq:tau_I-app}).
This suppression is supposed to
grow stronger  with the number of reflections; in other words, there is a further cost in amplitude for each trip through the
cavity ~\cite{Cardosorev}. The degree of suppression per trip is a factor that must be put in by hand.

Such a suppression in the amplitude of  GWs means that they are partially absorbed by the compact object. Let us recall  that total reflection is really no different than unhindered propagation through the object, as the waves would then be perceived as being totally reflected at $\;r=0\;$. Meanwhile, the intermediary  case of     partial absorption requires an extremely strong interaction. Such interactions would  be expected to cause dissipation rather than coherent motion (see below). They are also not particularly realistic since standard matter interacts extremely weakly with GWs; in fact, the whole Universe is transparent to them, and our working assumption is that the same would be true even for the exotic forms of matter which  might  be found inside of an ultracompact object. One could try to model the partial absorption with an ad-hoc modification of the gravitational potential, as done for example in \cite{yanbi}, but the justification for such an arbitrary manipulation  is lacking.

Nevertheless, as long as the absorption is not close to perfect, the spectral parameters of the echo modes  and their total energy will not be significantly modified,  as argued above. Recall in particular our argument that the absorption through the object's outer  surface  scales as $1/|\ln{\epsilon_i}|^2\;$ or, more precisely, as $1/\tau_i$. This means that the deviation from total reflection should scale similarly. In terms of a suppression coefficient $\gamma$, this
deviation would amount to $\;\gamma \sim 1-r_+/\tau\;$, where $\;\gamma=1\;$
translates into zero suppression. For GW150914, this would correspond to $\;1-\gamma\simeq 4\times 10^{-8}\;$.
Meaning that our bounds on the entropy can be expected to remain intact even when some absorption occurs.

\item
 The echo modes are also subject, in principle,  to intrinsic dissipation. The analysis to follow is premised on  the  assumption that such dissipation is negligible, which is consistent with total reflection as the echoing process requires essentially no interactions between the emitted GWs and the interior of the compact object.

\item

Formulating the full reflection of GWs from a surface is a rather daunting challenge and it is even unclear if a  consistent description exists for the case of rotating objects \cite{Price}. It is likely though that, for a slow-enough  rotation,  various approximations should be applicable  with a reasonable degree of accuracy.

\end{enumerate}

\section{Bound on entropy from gravitational-wave observations}

In this section,  following to some extent our own  methodology  from~\cite{collision}, we  derive possible bounds on the entropy for models of  BH alternatives.  The detection of echoes is currently under debate. Some groups claim that echoes do exist in the signals that were detected by  the LIGO/Virgo Collaboration~\cite{noncardoso,Abedi:2017isz,Conklin:2017lwb}, whereas the Collaboration itself concluded that the observational evidence for the existence of echoes is not significant enough to make such a claim~\cite{Westerweck:2017hus,Nielsen}. We
will adopt  a conservative standpoint and assume that GW signals from the remnant for GW150914 \cite{LIGO,LIGO1} are consistent with the QNMs of a BH. We will
then use the  fact that echoes were not detected by the LIGO/Virgo Collaboration   to see if one can place bounds on the entropy of the remnant of the merger.

We will be making  use of  Eqs.~\eqref{eq:omega_R-maggio} and~\eqref{eq:tau_I-maggio} (and also~\cite{Maggio:2017ivp,Cardosorev})
and only be considering the dominant $\;s=\ell=m= 2\;$ mode of
the gravitons. In what follows, a subscript of either $\rm{echo}$ or $i$  indicates a quantity that is relevant to the QN echo modes (the latter for the  $i^{\rm th}$ model),  whereas a subscript of $BH$ is reserved for the QNMs of a  classical BH.

\subsection{Gravitational wave spectra}
\label{sec:GW-spectra}

The QNM amplitude for the $i^{\rm th}$ model of exotic objects $A_i$, can be estimated by considering the total energy (and angular momentum) that the echoes carry away with them when they escape through the barrier.

There are two distinct sources of energy for the echoes. One is the initial energy that they have after the merger and  the other is the rotational energy of the BH. To understand the latter, notice that the frequency of the echo modes is related to the rotational frequency of the BH as in Eq.~(\ref{eq:omega-app}). The energy $\delta E$ and the angular momentum $\delta J$  that are depleted from the BH are related  by the first law of BH mechanics, $\;\delta E = \Omega \delta J\;$,
provided that the surface area does not change in the process (as would be the case when the process is adiabatic). The echoes' energy and angular momentum   and therefore their angular frequency all decrease because of their leakage through the gravitational potential barrier. Since the angular frequencies of the BH and echoes are correlated, it follows that the BH's energy and angular momentum,
as well as its  rotational energy, are  similarly all decreasing.

To estimate the initial energy of the echoes, we can use the fact, as predicted
by GR \cite{badri}, that roughly the same flux of energy is flowing inwards towards the surface of the compact object as flowing outwards in the form of standard prompt  GWs. For the case of GW150914, this is a flux of about 3 solar masses,
and so the initial energy of the echoes should be similar. However, we do expect that part of the initial inward flux energy will be emitted promptly and therefore only the trapped part of the energy will excite the echoes.

To estimate the rotational energy of the BH, we will use a non-relativistic approximation. The magnitude of the BH angular momentum is approximately given by
$
\;J \sim M r_+ v\;
$,
where $v$ is the BH rotational velocity. This leads to a  rotational kinetic energy  of the form
\be
E_\mathrm{rot} \;\sim\; \frac{1}{2} M v^2 \;\sim\; \frac{M \chi^2}{2(2+2\sqrt{1-\chi^2}-\chi^2)}\,,
\ee
where $\;\chi = a/M= J/M^2\;$ is the dimensionless spin parameter. Notice that
 the dependence of  the rotational energy  on $\chi$ is approximately quadratic.
Setting the  mass and spin parameters to match those of the   GW150914 remnant,
 $\;M=62.8\;M_\odot\;$ and  $\;\chi = 0.68\;$,  one then finds
that  $\;E_\mathrm{rot} \sim 4.8\; M_\odot\;$. This is a substantial fraction of the total energy of the BH.

The conclusion is that the total amount of energy (initial and rotational) which the echoes can carry and emit in the form of GWs is, in fact,  more than double the total amount of energy that can be emitted by a classical GR BH. This result is significant because the prospects for detection of such modes depends on the total energy emitted \cite{Flanagan:1997sx,Berti:2005ys}, as we now recall.

To obtain a  relationship between the amplitude of the echo modes and their total energy, we first consider the luminosity of the GWs,
\be
\frac{dE_{GW}}{dt}\; =\; \frac{1}{32\pi} D^2 \int d\Omega\, \langle \dot h_{\mu\nu} \dot h^{\mu\nu} \rangle\;.
\ee
Here, $D$ is the distance to the source of GWs, $h$ is the waveform in the transverse--traceless gauge, $d\Omega$ is an element of solid angle and angular brackets denote averaging over short wavelengths.  As an order-of-magnitude estimate, the integral in the above equation can be approximated as
\be
\int d\Omega\, \langle \dot h_{\mu\nu} \dot h^{\mu\nu} \rangle \;\sim\;  8\pi \dot h^2\;.
\ee

The amount of energy that is emitted in GWs for a duration of time $\Delta T$ can now be expressed as
\bea
\label{eq:E-rough}
E_{GW} &\sim & \frac{1}{32\pi} D^2 \, 8\pi \dot h^2\,  \Delta T  \;\sim\; \pi^2 D^2 f^2 |\tilde h|^2 \Delta T \;,
\eea
where $\;f=\omega/2\pi\;$ denotes a  spectral frequency
and  $\tilde h$ is the Fourier transform of $h$.
It follows that the ratio of the  energy which is  emitted in GWs by  standard GR processes to  the energy of the echoes  is given by
\be
\label{ratioE}
\frac{E_{BH}}{E_{\mathrm{echo}}}\;=\; \left(\frac{f_{BH}}{f_{\mathrm{echo}}}\right)^2 \frac{|\tilde h_{BH}|^2}{|\tilde h_{\mathrm{echo}}|^2} \frac{\Delta T_{BH}}{\Delta T_{\mathrm{echo}}}\;.
\ee

As will be shown below, one can estimate the signal-to-noise ratio (SNR) by estimating the energy of the GWs that the echoes emit. We, therefore, do not need to know explicitly the value of $|\tilde h_{\mathrm{echo}}|^2$. It is clear, however, that
$\;|\tilde h_{\mathrm{echo}}|^2 \ll |\tilde h_{BH}|^2\;$. This is because $\;\frac{\Delta T_{BH}}{\Delta T_{\mathrm{echo}}} \sim \frac{\tau_{BH}}{\tau_{\mathrm{echo}}}\sim 1/|\ln \epsilon|^2 \;$, while  $\left(\frac{f_{BH}}{f_{\mathrm{echo}}}\right)^2\;$ and $\frac{E_{BH}}{E_{\mathrm{echo}}}$ are parametrically of order unity. It then   follows that  $\;\frac{|\tilde h_{BH}|^2}{|\tilde h_{\mathrm{echo}}|^2} \sim |\ln \epsilon|^2\gg 1\; $.

Let us now recall  that the SNR $\rho$ for monochromatic GWs of frequency $f$ is
expressible as~\cite{Cutler:1997ta}
\be
\rho^2 \;=\; \frac{2}{S_n(f)} \int h(t)^2 \, dt\;,
\ee
with $S_n$ denoting the noise spectral density at $f$.
This expression can be approximated as
\be
\label{eq:SNR2-rough}
\rho^2 \;\sim \; \frac{2}{S_n(f)} h^2 \Delta T\;.
\ee

To estimate the amount of emitted GW energy in terms of the SNR, we
can combine Eqs.~\eqref{eq:E-rough} and~\eqref{eq:SNR2-rough}, which yields
\bea
\label{eq:E-in-rho}
E_{GW} & \sim & \frac{\pi^2}{2} D^2 f^2 \rho^2 S_n \nonumber \\
& \sim & 2.9 \, M_\odot \left( \frac{D}{430\mathrm{Mpc}} \right)^2 \left( \frac{f}{200\mathrm{Hz}} \right)^2 \nonumber \\
&&\times  \left( \frac{\rho}{24} \right)^2 \left( \frac{\sqrt{S_n}}{8 \times 10^{-23}\, \mathrm{Hz}^{-1/2}} \right)^2\,,
\eea
where GW150914 values have been incorporated into the lower of the two approximate equalities. Notice that this amount of emitted  energy  is in  excellent agreement with the amount $3 M_\odot$ as reported by the LIGO/Virgo Collaboration~\cite{LIGO}.

Alternatively, one can estimate the SNR in the echo modes from the fact that the maximum amount of additional energy that such modes can extract from the BH is
$E_\mathrm{rot}$. So that, by inverting Eq.~\eqref{eq:E-in-rho} and taking the sum of the initial energy for the echoes $2.9\; M_\odot$ and the rotational energy  $4.8\;M_\odot$ to give $\;E_{\mathrm{echo}}=7.7\;M_\odot\;$, one obtains
\bea
\label{eq:SNR-echo}
\rho_\mathrm{echo} &\sim& \frac{1}{\pi D f} \sqrt{\frac{2 E_\mathrm{echo}}{S_n}} \nonumber \\
&\sim & 39 \times \left(\frac{430\mathrm{Mpc}}{D} \right)  \left(\frac{200\mathrm{Hz}}{f}\right)  \nonumber \\
&&\times  \left( \frac{E_\mathrm{echo}}{7.7 M_\odot} \right)^{1/2} \frac{\sqrt{8 \times 10^{-23}\, \mathrm{Hz}^{-1/2}}}{\sqrt{S_n}}\;.
\eea
Thus, the echo modes would still have been detected, if present, provided that the search to detect them was carried over a long-enough period of time (see below).

\subsection{Current bounds from GW150914}

When a horizonless compact object is formed as a result of a merger, additional modes --- and echo modes in particular ---  can be expected as previously discussed.  Here, we are making the conservative assumption  that such modes are absent for GW150914.
As discussed at the beginning of the section, this premise  allows us to  place bounds on $\epsilon_i$  and, hence, on  the entropy of the remnant BH.

Let  us begin by considering  the  GW spectrum of the echo modes.
The Fourier waveform of the QNMs for a BH alternative can be expressed
as follows~\cite{Berti:2005ys,collision}:
\bea
\label{eq:htilde}
\tilde h(f) &=& e^{2\pi i f t_i} A_i \tau_i \frac{2f_i^2 Q_i \cos \phi_i- f_i (f_i-2 i f Q_i) \sin \phi_i}{f_i^2-4i f f_i Q_i + 4 (f_i^2-f^2)Q_i^2}\;,
\eea
for which  the square of the root-mean-square  magnitude goes as
\be
\label{eq:htildesquare}
|\tilde h|^2 \;=\; \frac{4 A_i^2 f_i^2 Q_i^4}{\pi ^2 \left\{16
   f^2 f_i^2 Q_i^2+\left[4 Q_i^2
   \left(f_i^2-f^2\right)+f_i^2\right]^2\right\}}\,,
\ee
where
$A_i$ is  the amplitude of the corresponding time-domain waveform, $f_i = \omega_i/2\pi$ is the frequency of the echo mode, $Q_i$ is defined by $\;Q_i= \pi f_i\tau_i\;$,  $t_i$ is the time delay of an echo mode relative to a typical BH mode and  $\phi_i$ is a  constant phase which will be set to zero for simplicity. It should be kept in mind  that, for the case of interest,  $\;|\ln{\epsilon_i}|\gg 1\;$, $\;f_i \propto  m\Omega\;$, $\;\tau_i \propto  |\ln{\epsilon_i}|^2\;$ and so $\;Q_i \gg 1\;$. This means that the width in the frequency domain of $|\tilde h|^2$ is small, making the emitted GW almost monochromatic. Notice that  the maximal  amplitude in the frequency domain, $\;|\tilde{h}(f_i)|^2 =\frac{1}{4} A_i^2 \tau_i^2\;$, depends on the decay time as well as the amplitude in the time domain $A_i$.

The amplitude of the echo mode $A_i$ is related to that of a BH mode $A_{BH}$ (given by Eq.~(21) of~\cite{collision}) as $\;A_i = \alpha_i A_{BH}\;$, where $\alpha_i <1$ is a constant that suppresses  the echo amplitude relative to the BH one. We determine this constant by requiring that the SNR of the echo mode to be $\;
\rho_i\sim 40\;$ as derived in Eq.~\eqref{eq:SNR-echo} and find that
$\;\alpha_i \sim 3 \times 10^{-3}\;$.

\begin{figure}[t]
\vspace{-3cm}
\centerline{\includegraphics[width=10.5cm]{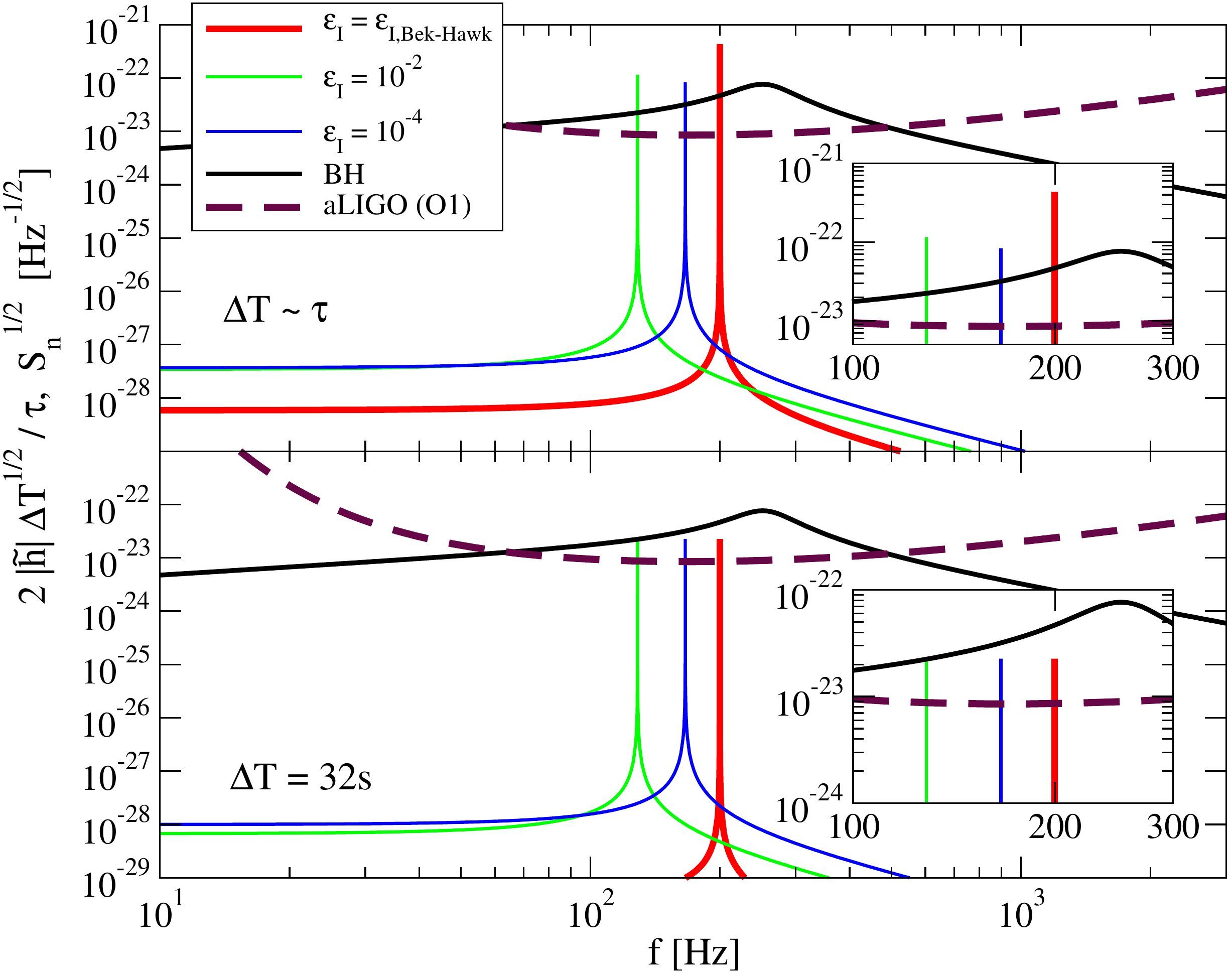}}
\caption{The GW spectra of additional QNMs for various $\epsilon_i$ (solid, colored), assuming that the remnant of GW150914 is an exotic alternative to a BH. We consider the analysis period of $\Delta T \sim \tau$ (top) and
$\Delta T = 32~{\rm s}$~\cite{Westerweck:2017hus,Nielsen} (bottom). For reference, we also present the spectrum for the QNMs of a GR BH  (black, solid) and the noise spectral density for aLIGO during its O1 run (brown, dashed). The insets zoom in the region close to peaks of the spectra. Note that establishing the absence of such modes with GW150914 would allow us to probe the Bekenstein--Hawking entropy (red, thick, solid). Here, we have scaled $|\tilde{h}|$ with $\Delta T^{1/2}/\tau$, as  both $1/\tau$ and $1/\Delta T$ are different from $f$.} \label{fig:spectrum-maggio} \end{figure}

The GW spectra for the echo modes are shown in the top panel of Fig.~\ref{fig:spectrum-maggio}, given that  the data-analysis time (or effective observation time) is at least comparable to the mode decay time $\tau_i$.  Also depicted is  the noise spectral density of Advanced LIGO (aLIGO) in its O1 run. Roughly speaking, the ratio between the GW  and  noise spectra corresponds to the SNR ratio of the putative additional modes. Although unlike, for instance, Fig.~13 of~\cite{Abbott:2018wiz}, we
are effectively displaying $2 |\tilde h|/ \sqrt{\tau_i}$ instead of $2 |\tilde h| \sqrt{f}$. This is because $1/\tau_i$ is much smaller than $f$ and the ratio between $2 |\tilde h|/ \sqrt{\tau_i}$ and $\sqrt{S_n}$ provides a closer estimate of the actual SNR.

Let us now comment on the suppression of the SNR due to the lack of sufficient
analysis time. The aforementioned SNR of $\sim 40$  is  only true for  data whose corresponding  time of analysis is larger than the echo damping time of
$\sim 10^4$~s; otherwise, the very narrow peaks in the spectra of  Fig.~\ref{fig:spectrum-maggio} cannot be resolved. Unfortunately, the LIGO/Virgo Collaboration has analyzed the data only up to 32~s after the merger (other parts of the signal were used for a time-slide analysis to estimate the noise)~\cite{Westerweck:2017hus}. This roughly leads to a reduction in the SNR of $\;\rho_i \sim \sqrt{32/10^4} \times 40 \sim 2\;$ (see the bottom panel of Fig.~\ref{fig:spectrum-maggio}). Thus, the current analysis cannot reveal the presence of the putative additional modes, as  one would need to analyze a longer portion of the data.

As can be seen in Fig.~\ref{fig:spectrum-maggio}, as well as in Eq.~\eqref{eq:omega_R-maggio}, the frequency of the additional modes increases with decreasing $\epsilon_i$ and reaches  a constant value of $\;2\Omega/2\pi \approx 200$~Hz for very small values of $\epsilon_i$. Because the limiting frequency is in the LIGO sensitivity band, it should be possible to place very stringent bounds on the existence of additional modes, assuming a longer time of analysis.

Let us now estimate the attainable bounds on $\epsilon_i$ using the upper panel of Fig.~\ref{fig:spectrum-maggio}. This clearly shows that aLIGO would have detected the additional echo modes if $\;10^{-40}\lesssim\epsilon_i \lesssim 0.01\;$. The lower bound comes from our prior belief that it does not make sense to consider $\epsilon_i$ smaller than $l_P/r_+\;$ (see below for more details). Therefore, if the absence of  additional modes would indeed be verified, as we expect, it would allow for the formal  exclusion of  this range of  $\epsilon_i$
values.   Meaning  that, formally, we would then be able to  deduce that either $\;S_i\gtrsim 10^{80}\;$ or $\;S_i\lesssim 10^4\;$.

Comparing our results to those found in \cite{Testa:2018bzd}, we find them to be consistent. For instance, the sharp peaks found in Fig.~1  are compatible with the sharp peaks in Fig.~1 of \cite{Testa:2018bzd} and, furthermore,  the finding in \cite{Testa:2018bzd} that aLIGO can detect echoes if the reflectivity coefficient $R$ is close to unity is consistent with our argument that the signal-to-noise ratio  for GW150914 echoes would have been about  $40$  had all the data been analyzed.

Values of $\;\epsilon_i \gtrsim 0.01\;$ are outside the domain of validity of our analysis  because, as previously emphasized, our framework requires $\;\epsilon_i \ll 1\;$.  So that, formally, the range $\;\epsilon_i \gtrsim 0.01\;$  cannot be  ruled out by our analysis and, therefore,  we cannot  argue conclusively against very small entropies. It is, however, highly unlikely that an astrophysical,  compact object ---  which is representing the final state of  collapsing matter ---  could
have an entropy   as low as $10^4$. Moreover, if $\epsilon_i$ is too
large, the inspiral part of the waveform becomes inconsistent with that of binary BHs due to their non-vanishing tidal deformabilities~\cite{Cardoso:2017cfl}.

On the other hand, since it does not make sense to consider $\epsilon_i$ any smaller than the effective limit of $l_P/r_+\;$, we are left with a ``range'' of $\;\epsilon_i \sim l_P/r_+\sim 10^{-40}\;$. This  range, as mentioned above, corresponds to an entropy of  $\;S_i \sim 10^{80}\;$,  a value which is of  the order of the Bekenstein--Hawking entropy for a BH of this same mass, $\;S_\mrm{Bek-Hawk} \simeq 3 \times 10^{80}\;$.

One might argue that the bound from the non-observation of the echoes should be interpreted as a bound on their amplitude or, equivalently, on their total energy  rather than on $\epsilon_i$.  But this is highly unlikely, as we now argue. To begin, let us recall that roughly the same flux of energy is flowing inwards towards the surface of the compact object as flowing outwards in the form of GWs \cite{badri}. Therefore, for the initial energy of the echos to be parametrically smaller than $E_{GW}$,  the reflection coefficient $R$ has to be parametrically smaller than unity. However, for any  regular, compact object without a horizon, no matter how exotic it might otherwise be, the coefficient $R$ has to be close to unity.~\footnote{The coefficient $R$ can be evaluated by estimating the reflection from the gravitational barrier near the surface of the object. This should be done for a momentum transfer that is twice the frequency of the waves.} This stands to reason because such a  relatively smaller-entropy object  should have a lower absorption rate than that  of a larger-entropy object like a GR BH. Furthermore, even when the reflection is small (but not parametrically  small), a substantial amount of the  the rotational energy and angular momentum of the compact object are shared with echoes. This process might be slower than expected if the reflection is small, but the relevant time scale  is still expected to be on the order of the
time for the round trip  (for the echoes to travel between the outer surface and the gravitational barrier)  and, therefore, considerably shorter than the decay time of the echoes. So that, even if for some reason the initial energy of the echoes was small, this  energy would be built up to a much larger value later
on.

To summarize, we expect that the GW150914 merger event most likely produced an object with an entropy on the order of the Bekenstein--Hawking value. A dedicated analysis should be able to verify our expectation.

\section{Conclusion}

We have proposed that BH mergers should be viewed as  cosmological scattering experiments which provide an opportunity to infer, or put bounds on, the entropy of  astrophysical BHs. Our main idea is that a lower bound on the entropy of a classical BH or any other exotic, ultracompact,  astrophysical object, can be deduced from a bound on the minimal (Tolman) redshift factor of waves emerging from the vicinity of its outer surface.  This minimal redshift factor can be, and already has been, constrained by observational data. More specifically, we have argued that the entropy of any ultracompact,  astrophysical object can be expected to scale with the inverse of the square of this minimal redshift factor.

Our argument closely follows logic that was first used by 't~Hooft  to show that the entropy of a BH does indeed agree with the Bekenstein--Hawking area law, subject to the assumption that the exterior metric is Schwarzschild regardless of the boundary conditions at the surface.  We have used the scaling relation between the entropy and the minimal redshift factor to show that a large range of entropy values could be excluded by data from the  GW150914 event. If this analysis is done for a long-enough period of time,  we would expect it to show that astrophysical BHs most likely have an extremely large entropy; on the order of the Bekenstein--Hawking entropy. This would be  $\;S\sim 10^{80}\;$ for the remnant  of GW150914.

Formally, our analysis cannot exclude very low entropy values on  the order of $10^4$ or smaller. This range is  not excluded because our framework is only valid for values of the minimal redshift factor that are parametrically small. It is, however, quite unlikely that an astrophysical, ultracompact object
which represents the final state of  collapsing matter  could have an entropy   as low as $10^4$.

Our bounds are based on the  assumption that  any given exotic alternative has a reflecting outer surface. Nevertheless, we have also argued that the degree of reflection is not so important for the validity of the bounds. Even  if the reflection is not complete, the bounds would still be  strong because, for exotic
objects with small-enough values of $\epsilon_i$,
the frequencies of the additional modes  lie in the LIGO sensitivity band and their total energy is larger than those of the  standard BH modes.

To reach a precise quantitative lower bound on the remnant entropy, a more sophisticated analysis is required, as in \cite{Westerweck:2017hus,Nielsen,Weinstein},  but for a longer duration.  Besides using actual data, the analysis should account for  the possibility of some absorption in a quantitatively precise way and likewise for the possible instabilities. However, our results do indicate that such an analysis would  also yield bounds of the order of the Bekenstein--Hawking entropy.~\footnote{The uncertainty principle of quantum gravity may set a fundamental resolution limit to  GW observations that is well above the Planckian scale~\cite{Addazi:2018uhd}. Our estimate is different from~\cite{Addazi:2018uhd} as we do not \emph{measure} any Planckian effect and the lower bound on the entropy partially comes from a theoretical consideration.} Future planned third-generation interferometers such as the Einstein Telescope \cite{Hild:2010id} and the Cosmic Explorer \cite{Evans:2016mbw} would be able to provide much more accurate results.

\section*{Acknowledgments}

We would like to thank Bruce Allen, Vitor cardoso, Badri Krishnan, Alex Nielsen, Paolo Pani, Jing Ren and Nicolas Yunes for useful discussions and comments on the manuscript. Special thanks to Collin Capano for detecting an error in our estimate of the SNR in a previous version and for discussing the details of the LIGO/VIRGO analysis.
The research of AJMM received support from an NRF Incentive Funding Grant 85353
and an NRF Competitive Programme Grant 93595. The research of RB was supported by the Israel Science Foundation grant no. 1294/16. AJMM thanks Ben Gurion University for their  hospitality during his visit.
KY acknowledges support from NSF Award PHY-1806776.
KY would like to also acknowledge networking support by the COST Action GWverse CA16104.


\begin{thebibliography}{99}




\bibitem{Sunny}
 N.~Itzhaki,
  ``Is the black hole complementarity principle really necessary?,''
 arXiv:hep-th/9607028.

\bibitem{Mathur1}
  S.~D.~Mathur,
  ``What Exactly is the Information Paradox?,''
  Lect.\ Notes Phys.\  {\bf 769}, 3 (2009)
  [arXiv:0803.2030 [hep-th]].




\bibitem{Braun}
 S.~L.~Braunstein, S.~Pirandola and K.~Zyczkowski,
  ``Entangled black holes as ciphers of hidden information,''
  Physical Review Letters 110, {\bf 101301} (2013)
  [arXiv:0907.1190 [quant-ph]].



\bibitem{Mathur2}
  S.~D.~Mathur,
``The Information paradox: A Pedagogical introduction,''
  Class.\ Quant.\ Grav.\  {\bf 26}, 224001 (2009)
  [arXiv:0909.1038 [hep-th]].




\bibitem{AMPS}
 A.~Almheiri, D.~Marolf, J.~Polchinski and J.~Sully,
  ``Black Holes: Complementarity or Firewalls?,''
  JHEP {\bf 1302}, 062 (2013)
  [arXiv:1207.3123 [hep-th]].







\bibitem{MP}
D.~Marolf and J.~Polchinski,
  ``Gauge/Gravity Duality and the Black Hole Interior,''
  Phys.\ Rev.\ Lett.\  {\bf 111}, 171301 (2013)
  [arXiv:1307.4706 [hep-th]].




\bibitem{MathurCP}
  S.~D.~Mathur,
  ``Resolving the black hole causality paradox,''
  arXiv:1703.03042 [hep-th].




\bibitem{otherfuzzball}
 K.~Skenderis and M.~Taylor,
  ``The fuzzball proposal for black holes,''
  Phys.\ Rept.\  {\bf 467}, 117 (2008)
  [arXiv:0804.0552 [hep-th]].




\bibitem{Cardoso}
   V.~Cardoso, E.~Franzin and P.~Pani,
  ``Is the gravitational-wave ringdown a probe of the event horizon?,''
  Phys.\ Rev.\ Lett.\  {\bf 116}, no. 17, 171101 (2016)
  Erratum: [Phys.\ Rev.\ Lett.\  {\bf 117}, no. 8, 089902 (2016)]
  [arXiv:1602.07309 [gr-qc]].



\bibitem{MM}
  P.~O.~Mazur and E.~Mottola,
  ``Gravitational condensate stars: An alternative to black holes,''
  gr-qc/0109035.

\bibitem{bubble}
  U.~H.~Danielsson, G.~Dibitetto and S.~Giri,
  ``Black holes as bubbles of AdS,''
  JHEP {\bf 1710}, 171 (2017)
  [arXiv:1705.10172 [hep-th]].

\bibitem{bozo}
  D.~F.~Torres, S.~Capozziello and G.~Lambiase,
  ``A Supermassive scalar star at the galactic center?,''
  Phys.\ Rev.\ D {\bf 62}, 104012 (2000)
  [astro-ph/0004064].



\bibitem{Dvali}
  G.~Dvali and C.~Gomez,
  ``Black Hole's Quantum N-Portrait,''
  Fortsch.\ Phys.\  {\bf 61}, 742 (2013)
  [arXiv:1112.3359 [hep-th]].

\bibitem{Bowers:1974tgi}
  R.~L.~Bowers and E.~P.~T.~Liang,
  ``Anisotropic Spheres in General Relativity,''
  Astrophys.\ J.\  {\bf 188}, 657 (1974).

\bibitem{Yagi:2015upa}
  K.~Yagi and N.~Yunes,
  ``Relating follicly-challenged compact stars to bald black holes: A link between two no-hair properties,''
  Phys.\ Rev.\ D {\bf 91}, no. 10, 103003 (2015)
  [arXiv:1502.04131 [gr-qc]].

  \bibitem{Yagi:2016ejg}
  K.~Yagi and N.~Yunes,
  ``I-Love-Q Relations: From Compact Stars to Black Holes,''
  Class.\ Quant.\ Grav.\  {\bf 33}, no. 9, 095005 (2016)
  [arXiv:1601.02171 [gr-qc]].

\bibitem{Raposo:2018rjn}
  G.~Raposo, P.~Pani, M.~Bezares, C.~Palenzuela and V.~Cardoso,
  ``Anisotropic stars as ultracompact objects in General Relativity,''
  arXiv:1811.07917 [gr-qc].

\bibitem{strungout}
R. Brustein and A. J. M. Medved,
  ``Black holes as collapsed polymers,''
  Fortsch.\ Phys.\  {\bf 65}, 0114 (2017)
  [arXiv:1602.07706 [hep-th]].



\bibitem{Cardosorev}
   V.~Cardoso and P.~Pani,
  ``Tests for the existence of horizons through gravitational wave echoes,''
  Nat.\ Astron.\  {\bf 1}, 586 (2017)
  [arXiv:1709.01525 [gr-qc]];
  ``The observational evidence for horizons: from echoes to precision gravitational-wave physics,''
  arXiv:1707.03021 [gr-qc].


\bibitem{bek}
  J.~D.~Bekenstein,
  ``Do we understand black hole entropy?,''
  gr-qc/9409015.

\bibitem{brick}
 G.  't  Hooft,
``On the quantum structure of a black hole,'' Nucl. Phys. B
{\bf 256}, 727 (1985).



\bibitem{LIGO}
B. P. Abbott {\em et al}. [LIGO Scientific and Virgo Collaborations], ``Observation of Gravitational Waves from a Binary Black Hole Merger,''
Phys. Rev. Lett. {\bf 116}, no. 6, 061102 (2016) [arXiv:1602.03837 [gr-qc]].


\bibitem{TheLIGOScientific}
  B.~P.~Abbott {\it et al.} [LIGO Scientific and Virgo Collaborations],
  ``Binary Black Hole Mergers in the first Advanced LIGO Observing Run,''
  Phys.\ Rev.\ X {\bf 6}, no. 4, 041015 (2016)
  [arXiv:1606.04856 [gr-qc]].




\bibitem{LIGO1}
  B.~P.~Abbott {\it et al.} [LIGO Scientific and Virgo Collaborations],
  ``Tests of general relativity with GW150914,''
  Phys.\ Rev.\ Lett.\  {\bf 116}, no. 22, 221101 (2016)
  [arXiv:1602.03841 [gr-qc]].



\bibitem{Konoplya:2016pmh}
  R.~Konoplya and A.~Zhidenko,
  ``Detection of gravitational waves from black holes: Is there a window for alternative theories?,''
  Phys.\ Lett.\ B {\bf 756}, 350 (2016)
  [arXiv:1602.04738 [gr-qc]].


\bibitem{pretorius}
  N.~Yunes, K.~Yagi and F.~Pretorius,
  ``Theoretical Physics Implications of the Binary Black-Hole Mergers GW150914 and GW151226,''
  Phys.\ Rev.\ D {\bf 94}, no. 8, 084002 (2016)
  [arXiv:1603.08955 [gr-qc]].


\bibitem{AKK}
N. Andersson, Y. Kojima and K. D.  Kokkotas,
``On the Oscillation Spectra of Ultracompact Stars: an Extensive Survey of Gravitational-Wave Modes,''
Astrophys. J. {\bf 462}, 855 (1996).

\bibitem{spinny}
  R.~Brustein, A.~J.~M.~Medved and K.~Yagi,
  ``Discovering the interior of black holes,''
  Phys.\ Rev.\ D {\bf 96}, no. 12, 124021 (2017)
  [arXiv:1701.07444 [gr-qc]].

\bibitem{collision}
  R.~Brustein, A.~J.~M.~Medved and K.~Yagi,
  ``When black holes collide: Probing the interior composition by the spectrum of ringdown modes and emitted gravitational waves,''
  Phys.\ Rev.\ D {\bf 96}, no. 6, 064033 (2017)
  [arXiv:1704.05789 [gr-qc]].

\bibitem{moreCardoso}
  V.~Cardoso, S.~Hopper, C.~F.~B.~Macedo, C.~Palenzuela and P.~Pani,
  ``Gravitational-wave signatures of exotic compact objects and of quantum corrections at the horizon scale,''
  Phys.\ Rev.\ D {\bf 94}, no. 8, 084031 (2016)
  [arXiv:1608.08637 [gr-qc]].

\bibitem{noncardoso}
  J.~Abedi, H.~Dykaar and N.~Afshordi,
  ``Echoes from the Abyss: Evidence for Planck-scale structure at black hole horizons,''
  arXiv:1612.00266 [gr-qc];
  ``Echoes from the Abyss: The Holiday Edition!,''
  arXiv:1701.03485 [gr-qc].

\bibitem{Conklin:2017lwb}
  R.~S.~Conklin, B.~Holdom and J.~Ren,
  ``Gravitational wave echoes through new windows,''
  Phys.\ Rev.\ D {\bf 98}, no. 4, 044021 (2018)
  doi:10.1103/PhysRevD.98.044021
  [arXiv:1712.06517 [gr-qc]].

\bibitem{Price}
  R.~H.~Price and G.~Khanna,
  ``Gravitational wave sources: reflections and echoes,''
  Class.\ Quant.\ Grav.\  {\bf 34}, no. 22, 225005 (2017)
  [arXiv:1702.04833 [gr-qc]].

\bibitem{Westerweck:2017hus}
  J.~Westerweck {\it et al.},
  ``Low significance of evidence for black hole echoes in gravitational wave data,''
  Phys.\ Rev.\ D {\bf 97}, no. 12, 124037 (2018)
  [arXiv:1712.09966 [gr-qc]].

  \bibitem{Pani:2010jz}
  P.~Pani, E.~Barausse, E.~Berti and V.~Cardoso,
  ``Gravitational instabilities of superspinars,''
  Phys.\ Rev.\ D {\bf 82}, 044009 (2010)
  [arXiv:1006.1863 [gr-qc]].

\bibitem{Cardoso:2014sna}
  V.~Cardoso, L.~C.~B.~Crispino, C.~F.~B.~Macedo, H.~Okawa and P.~Pani,
  ``Light rings as observational evidence for event horizons: long-lived modes, ergoregions and nonlinear instabilities of ultracompact objects,''
  Phys.\ Rev.\ D {\bf 90}, no. 4, 044069 (2014)
  [arXiv:1406.5510 [gr-qc]].

  \bibitem{Maggio:2017ivp}
  E.~Maggio, P.~Pani and V.~Ferrari,
  ``Exotic Compact Objects and How to Quench their Ergoregion Instability,''
  Phys.\ Rev.\ D {\bf 96}, no. 10, 104047 (2017)
  [arXiv:1703.03696 [gr-qc]].



\bibitem{Barausse:2018vdb}
  E.~Barausse, R.~Brito, V.~Cardoso, I.~Dvorkin and P.~Pani,
  ``The stochastic gravitational-wave background in the absence of horizons,''
  arXiv:1805.08229 [gr-qc].


\bibitem{yanbi}
  S.~M.~Du and Y.~Chen,
  ``Searching for near-horizon quantum structures in the binary black-hole stochastic gravitational-wave background,''
  Phys.\ Rev.\ Lett.\  {\bf 121}, no. 5, 051105 (2018)
  [arXiv:1803.10947 [gr-qc]].





  \bibitem{Maggio:2018ivz}
  E.~Maggio, V.~Cardoso, S.~R.~Dolan and P.~Pani,
  ``Ergoregion instability of exotic compact objects: electromagnetic and gravitational perturbations and the role of absorption,''
  arXiv:1807.08840 [gr-qc].


\bibitem{MM2}
  P.~O.~Mazur and E.~Mottola,
  ``Gravitational vacuum condensate stars,''
  Proc.\ Nat.\ Acad.\ Sci.\  {\bf 101}, 9545 (2004)
  [gr-qc/0407075].


\bibitem{RBKupf}
  R.~Brustein and J.~Kupferman,
  ``Black hole entropy divergence and the uncertainty principle,''
  Phys.\ Rev.\ D {\bf 83}, 124014 (2011)
  [arXiv:1010.4157 [hep-th]].

\bibitem{origin}
  R.~Brustein,
  ``Origin of the blackhole information paradox,''
  Fortsch.\ Phys.\  {\bf 62}, 255 (2014)
  [arXiv:1209.2686 [hep-th]].

\bibitem{Brustein:2013xga}
  R.~Brustein and A.~J.~M.~Medved,
  ``Semiclassical black holes expose forbidden charges and censor divergent densities,''
  JHEP {\bf 1309}, 108 (2013)
  [arXiv:1302.6086 [hep-th]].



\bibitem{Oppenheim:2002kx}
  J.~Oppenheim,
  ``Thermodynamics with long-range interactions: From Ising models to black holes,''
  Phys.\ Rev.\ E {\bf 68}, 016108 (2003)
  [gr-qc/0212066].

  \bibitem{Alexander:2018wxr}
  S.~H.~Alexander, K.~Yagi and N.~Yunes,
  ``An Entropy-Area Law for Neutron Stars Near the Black Hole Threshold,''
  arXiv:1810.01313 [gr-qc].

\bibitem{RBYarom}
  R.~Brustein, M.~B.~Einhorn and A.~Yarom,
  ``Entanglement interpretation of black hole entropy in string theory,''
  JHEP {\bf 0601}, 098 (2006)
  [hep-th/0508217].



\bibitem{Dvali2}
G.~Dvali,
  ``Black Holes and Large N Species Solution to the Hierarchy Problem,''
  Fortsch.\ Phys.\  {\bf 58}, 528 (2010)
  [arXiv:0706.2050 [hep-th]].

\bibitem{BDV}
  R.~Brustein, G.~Dvali and G.~Veneziano,
  ``A Bound on the effective gravitational coupling from semiclassical black holes,''
  JHEP {\bf 0910}, 085 (2009)
  [arXiv:0907.5516 [hep-th]].

\bibitem{brickrot}
  J.~l.~Jing and M.~L.~Yan,
  ``Quantum entropy of the Kerr black hole arising from gravitational perturbation,''
  Phys.\ Rev.\ D {\bf 64}, 064015 (2001)
  [gr-qc/0104054].


\bibitem{Holdom:2016nek}
  B.~Holdom and J.~Ren,
  ``Not quite a black hole,''
  Phys.\ Rev.\ D {\bf 95}, no. 8, 084034 (2017)
  [arXiv:1612.04889 [gr-qc]].

\bibitem{QNMBH}
  K.~D.~Kokkotas and B.~G.~Schmidt,
  ``Quasinormal modes of stars and black holes,''
  Living Rev.\ Rel.\  {\bf 2}, 2 (1999)
  [gr-qc/9909058].

\bibitem{QNMBH2}
H.-P. Nollert,
``Quasinormal modes: the characteristic `sound' of black holes and
neutron stars,''
Class. Quant. Grav. {\bf 16}, R159 (1999).





\bibitem{Konoplya:2011qq}
   R.~A.~Konoplya and A.~Zhidenko,
   ``Quasinormal modes of black holes: From astrophysics to string theory,''
   Rev.\ Mod.\ Phys.\  {\bf 83}, 793 (2011)
   [arXiv:1102.4014 [gr-qc]].


\bibitem{Starobinsky}
  A.~A.~Starobinsky,
  ``Amplification of waves reflected from a rotating "black hole",''
  Sov.\ Phys.\ JETP {\bf 37}, no. 1, 28 (1973)
  [Zh.\ Eksp.\ Teor.\ Fiz.\  {\bf 64}, 48 (1973)].

\bibitem{Vilenkin}
  A.~Vilenkin,
  ``Exponential Amplification of Waves in the Gravitational Field of Ultrarelativistic Rotating Body,''
  Phys.\ Lett.\  {\bf 78B}, 301 (1978).

\bibitem{LOM}
L. Lindblom, B. J. Owen and S. M. Morsink, ``Gravitational radiation
instability in hot young neutron stars,''  Phys. Rev. Lett. {\bf 80}, 4843
(1998) [gr-qc/9803053].





\bibitem{ridethewave}
  R.~Brustein and A.~J.~M.~Medved,
  ``Quantum hair of black holes out of equilibrium,''
  Phys.\ Rev.\ D {\bf 97}, no. 4, 044035 (2018)
  [arXiv:1709.03566 [hep-th]].


\bibitem{Testa:2018bzd}
  A.~Testa and P.~Pani,
  ``Analytical template for gravitational-wave echoes: signal characterization and prospects of detection with current and future interferometers,''
  Phys.\ Rev.\ D {\bf 98}, no. 4, 044018 (2018)
  doi:10.1103/PhysRevD.98.044018
  [arXiv:1806.04253 [gr-qc]].

\bibitem{PPani}
  P.~Pani,
  ``I-Love-Q relations for gravastars and the approach to the black-hole limit,''
  Phys.\ Rev.\ D {\bf 92}, no. 12, 124030 (2015)
  Erratum: [Phys.\ Rev.\ D {\bf 95}, no. 4, 049902 (2017)]
  [arXiv:1506.06050 [gr-qc]].



\bibitem{Abedi:2017isz}
  J.~Abedi, H.~Dykaar and N.~Afshordi,
  ``Echoes from the Abyss: The Holiday Edition!,''
  arXiv:1701.03485 [gr-qc].


\bibitem{Nielsen}
  A.~B.~Nielsen, C.~D.~Capano and J.~Westerweck,
  ``Parameter estimation for black hole echo signals and their statistical significance,''
  arXiv:1811.04904 [gr-qc].

\bibitem{badri}
  A.~Gupta, B.~Krishnan, A.~Nielsen and E.~Schnetter,
  ``Dynamics of marginally trapped surfaces in a binary black hole merger: Growth and approach to equilibrium,''
  Phys.\ Rev.\ D {\bf 97}, no. 8, 084028 (2018)
  [arXiv:1801.07048 [gr-qc]].

\bibitem{Flanagan:1997sx}
  E.~E.~Flanagan and S.~A.~Hughes,
  ``Measuring gravitational waves from binary black hole coalescences: 1. Signal-to-noise for inspiral, merger, and ringdown,''
  Phys.\ Rev.\ D {\bf 57}, 4535 (1998)
  [gr-qc/9701039].



\bibitem{Berti:2005ys}
  E.~Berti, V.~Cardoso and C.~M.~Will,
 ``On gravitational-wave spectroscopy of massive black holes with the space interferometer LISA,''
  Phys.\ Rev.\ D {\bf 73}, 064030 (2006)
 doi:10.1103/PhysRevD.73.064030
  [gr-qc/0512160].

\bibitem{Cutler:1997ta}
  C.~Cutler,
  ``Angular resolution of the LISA gravitational wave detector,''
  Phys.\ Rev.\ D {\bf 57}, 7089 (1998)
  [gr-qc/9703068].


\bibitem{Abbott:2018wiz}
  B.~P.~Abbott {\it et al.} [LIGO Scientific and Virgo Collaborations],
  ``Properties of the binary neutron star merger GW170817,''
  arXiv:1805.11579 [gr-qc].

\bibitem{Cardoso:2017cfl}
  V.~Cardoso, E.~Franzin, A.~Maselli, P.~Pani and G.~Raposo,
  ``Testing strong-field gravity with tidal Love numbers,''
  Phys.\ Rev.\ D {\bf 95}, no. 8, 084014 (2017)
  Addendum: [Phys.\ Rev.\ D {\bf 95}, no. 8, 089901 (2017)]
  [arXiv:1701.01116 [gr-qc]].




\bibitem{Weinstein}
  R.~K.~L.~Lo, T.~G.~F.~Li and A.~J.~Weinstein,
  ``Template-based Gravitational-Wave Echoes Search Using Bayesian Model Selection,''
  arXiv:1811.07431 [gr-qc].


  \bibitem{Addazi:2018uhd}
  A.~Addazi, A.~Marciano and N.~Yunes,
  ``Can we probe Planckian corrections at the horizon scale with gravitational waves?,''
  arXiv:1810.10417 [gr-qc].





\bibitem{Hild:2010id}
  S.~Hild {\it et al.},
  ``Sensitivity Studies for Third-Generation Gravitational Wave Observatories,''
  Class.\ Quant.\ Grav.\  {\bf 28}, 094013 (2011)
  [arXiv:1012.0908 [gr-qc]].


\bibitem{Evans:2016mbw}
  B.~P.~Abbott {\it et al.} [LIGO Scientific Collaboration],
  ``Exploring the Sensitivity of Next Generation Gravitational Wave Detectors,''
  Class.\ Quant.\ Grav.\  {\bf 34}, no. 4, 044001 (2017)
  [arXiv:1607.08697 [astro-ph.IM]].



\end{thebibliography}
\end{document}